\documentclass[twocolumn,prc,floatfix,showpacs,preprintnumbers,nofootinbib,%
superscriptaddress]{revtex4}
\usepackage{mathrsfs}
\usepackage{amssymb}
\usepackage{amsmath}
\usepackage{graphicx}

\usepackage{xcolor}
\definecolor{lcolor}{rgb}{0.5,0,0}
\definecolor{citcolor}{rgb}{0,0.3,0.0}

\usepackage[breaklinks,colorlinks,urlcolor=blue,citecolor=citcolor,linkcolor=lcolor]{hyperref}
\usepackage{mciteplus}

\def\be{\begin{equation}}
\def\ee{\end{equation}}
\def\bea{\begin{eqnarray}}
\def\eea{\end{eqnarray}}
\def\gsim{ \,\, \vcenter{\hbox{$\buildrel{\displaystyle >}\over\sim$}}
 \,\,}

\newcommand{\xt}{{\mathbf{x}_T}}

\newcommand{\yt}{{\mathbf{y}_T}}

\newcommand{\ut}{{\mathbf{u}_T}}
\newcommand{\vt}{{\mathbf{v}_T}}
\newcommand{\x}{{\mathbf{x}}}
\newcommand{\y}{{\mathbf{y}}}

\newcommand{\itt}{\mathbf{i}_T}
\newcommand{\jt}{\mathbf{j}_T}
\newcommand{\ktt}{k_T} % scalar
\newcommand{\nabt}{\boldsymbol{\nabla}_T}
\newcommand{\At}{\mathbf{A}_T}

\newcommand{\Avec}{\mathbf{A}}

\newcommand{\ud}{\, \mathrm{d}}

\newcommand{\tr}{\, \mathrm{Tr} \, }

\newcommand{\nc}{{N_\mathrm{c}}}

\newcommand{\nr}[1]{(\ref{#1})}

\newcommand{\qs}{Q_\mathrm{s}}

\newcommand{\lqcd}{\Lambda_{\mathrm{QCD}}}
\newcommand{\as}{\alpha_{\mathrm{s}}}

\newcommand{\fig}{Fig.~}
\newcommand{\figs}{Figs.~}
\newcommand{\eq}{Eq.~}

\begin{document}

\author{A. Dumitru}
\affiliation{Department of Natural Sciences, Baruch College, New York, NY 10010,
 USA}

\author{T. Lappi}
\affiliation{
Department of Physics, %
 P.O. Box 35, 40014 University of Jyv\"askyl\"a, Finland
}
\affiliation{
Helsinki Institute of Physics, P.O. Box 64, 00014 University of Helsinki,
Finland
}

\author{Y. Nara}
\affiliation{
Akita International University, Yuwa, Akita-city 010-1292, Japan
}

\title{
Structure of longitudinal chromomagnetic fields in high energy collisions
}


\preprint{}

\begin{abstract}
We compute expectation values of spatial Wilson loops in the forward
light cone of high-energy collisions. We consider ensembles of gauge
field configurations generated from a classical Gaussian effective
action as well as solutions of high-energy renormalization group
evolution with fixed and running coupling. The initial fields
correspond to a color field condensate exhibiting domain-like
structure over distance scales of order the saturation scale. At later
times universal scaling emerges at large distances for all ensembles,
with a nontrivial critical exponent. Finally, we compare the results
for the Wilson loop to the two-point correlator of magnetic fields.
\end{abstract}

\maketitle

%%%%%%%%%%%%%%%%%%%%%%%%%
\section{Introduction}
%%%%%%%%%%%%%%%%%%%%%%%%%%

Heavy ion collisions at high energies involve non-linear dynamics of
strong QCD color fields~\cite{Mueller:1999wm}. These soft fields
correspond to gluons with light-cone momentum fractions $x\ll1$,
which can be described in the ``Color Glass Condensate''
(CGC) framework.
Because of the high gluon occupation number the gluon field
can be determined
from the classical Yang-Mills equations with a static current on the
light cone~\cite{McLerran:1994ni,*McLerran:1994ka,*McLerran:1994vd}. It
consists of gluons with a transverse momentum on the order of the
density of valence charges per unit transverse area,
$\qs^2$~\cite{JalilianMarian:1996xn}. Parametrically, the saturation
momentum scale $\qs$ separates the regime of non-linear color field
interactions from the perturbative (linear) regime. It is commonly
defined using a two-point function of electric Wilson lines, the
``dipole scattering amplitude''  evaluated in the field
of a single hadron or nucleus~\cite{Kovchegov:1998bi} as described below.

Before the collision the individual fields of projectile and target
are two dimensional pure gauges; in light cone gauge,
\be \label{eq:alphai} 
\alpha^i_m = \frac{i}{g} \, V_m  \, \partial^i V_m^\dagger
\ee
where $m=1,\, 2$ labels the projectile and target, respectively. Here 
$V_m$ are light-like SU($\nc$) Wilson lines,
 which correspond to the
eikonal phase of a high energy projectile passing through the
classical field shockwave~\cite{Balitsky:1995ub,Buchmuller:1995mr}.

The field in the forward light cone after the collision
 up to the formation of a thermalized
plasma is commonly called the ``glasma''~\cite{Lappi:2006fp}.
Immediately after the collision
longitudinal chromo-electric and magnetic fields $E_z,~B_z\sim 1/g$
dominate~\cite{Kharzeev:2001ev,*Fries:2006pv,Lappi:2006fp}. They
fluctuate according to the random local color charge densities of the
valence sources.
The magnitude of the 
color charge fluctuations is related to the saturation
scale $\qs^2$.
The transverse gauge potential
at proper time $\tau\equiv\sqrt{t^2-z^2}\to 0$, 
is given by~\cite{Kovner:1995ja}
\be \label{eq:alpha1+alpha2}
A^i=\alpha_1^i + \alpha_2^i~.
\ee
Note that while the fields of the individual projectiles $\alpha_m^i$ 
are pure gauges, for a non-Abelian gauge
theory $A^i$ is not. Hence, spatial Wilson loops evaluated in the
field $A^i$ are not equal to 1. The field at later times
is then obtained from the classical Yang-Mills equations of motion, which
can be solved either analytically in an expansion in the
field strength~\cite{Kovner:1995ja,Blaizot:2008yb} or numerically on a
lattice~\cite{Krasnitz:1998ns,Krasnitz:2001qu,*Krasnitz:2003jw,*Lappi:2003bi,Lappi:2007ku}.
The Wilson loop, and the magnetic field correlator, provide
an explicitly gauge-invariant method to study the 
nonperturbative dynamics of these fields, complementary
to studies of the gluon spectrum~\cite{Lappi:2011ju}.

Spatial Wilson loops at very early times $\tau$ have recently been
studied numerically in Ref.~\cite{Dumitru:2013koh}, using the MV
model~\cite{McLerran:1994ni,*McLerran:1994ka,*McLerran:1994vd} for the
colliding color charge sheets.  It was observed that the loops
effectively satisfy area law scaling for radii $\gsim
1/\qs$, up to a few times this scale. Furthermore,
Ref.~\cite{Dumitru:2013wca} found that two-point correlators of $B_z$
over distances $\lesssim 1/\qs$ correspond to two dimensional
screened propagators with a magnetic screening mass a few times
$\qs$. This indicates that the initial fields exhibit {\em structure}
such that magnetic flux does not spread uniformly over the transverse
plane (like in a Coulomb phase) but instead is concentrated in small
domains.

The present paper extends this previous work as follows. We perform lattice
measurements of spatial Wilson loops over a much broader range of
radii to analyze their behavior at short ($R\ll1/\qs$) and long
($R\gg1/\qs$) distances. We also implement the so-called 
JIMWLK~\cite{
JalilianMarian:1996xn,Jalilian-Marian:1997jx,*Jalilian-Marian:1997gr,%
*Jalilian-Marian:1997dw,*JalilianMarian:1998cb,*Iancu:2000hn,%
*Iancu:2001md,*Ferreiro:2001qy,*Iancu:2001ad,Weigert:2000gi,*Mueller:2001uk}
high-energy functional renormalization group evolution which resums
observables to all orders in $\as \log(1/x)$. High-energy
evolution  modifies the classical ensemble of gauge field
configurations~\nr{eq:V_rho}, \nr{eq:S2} to account for nearly
boost invariant quantum fluctuations at rapidities far from the
sources. Finally, we also solve the Yang-Mills equations in the
forward light cone to study the time evolution of magnetic flux
loops.

The calculation of the initial conditions and the numerical solution
of the classical boost-invariant\footnote{The YM equations are solved
  in terms of the coordinates $\tau=\sqrt{t^2-z^2}$,
  $\eta=\frac{1}{2}\ln \frac{t+z}{t-z}$ and $\xt$; hence $\ud
  s^2=\ud\tau^2-\tau^2\ud\eta^2-\ud\xt^2 $.} Yang-Mills fields in the
initial stages of a heavy ion collision have been documented in
the references given below, so here we will only describe
them very briefly in Sec.~\ref{sec:numerics} before moving on
to show our results in Secs.~\ref{sec:wloop} and ~\ref{sec:bbcorr}.

%%%%%%%%%%%%%%%%%%%%%%%%%
\section{Lattice implementation}\label{sec:numerics}
%%%%%%%%%%%%%%%%%%%%%%%%%%

We work on a two dimensional square lattice of
$N_\perp^2$ points with periodic boundary conditions and consider
color sources that fill the whole transverse plane. The lattice
spacing is denoted as $a$, thus the area of the lattice in physical
units is $L^2=N_\perp^2a^2$.  The calculations are performed for
$\nc=3$ colors.  In this work we only consider symmetric collisions,
where the color charges of both colliding nuclei are taken from the
same probability distribution.

In this work we compare three different
initial conditions for the classical Yang-Mills equations:
 the classical MV model~\nr{eq:S2} as well as
fixed and running coupling JIMWLK evolution. We define the saturation
scale $\qs(Y)$ at rapidity $Y$ through the expectation value of the
dipole operator as
\begin{equation}\label{eq:defqs}
\frac{1}{\nc} \left\langle \tr V^\dag(\xt) V(\yt) 
\right\rangle_{Y,|\xt-\yt|=\sqrt{2}/\qs}
= e^{-1/2}.
\end{equation}
Throughout this paper we shall use $\qs$ defined in this way from the
light-like Wilson lines $V(\xt)$ in the fundamental representation.
The saturation scale is the only scale in the problem and we attempt
to construct the various initial conditions in such a way that the
value of $\qs a$ is similar, to ensure a similar dependence on
discretization effects.

In the MV model the Wilson lines are obtained from a classical color
charge density $\rho$ as
\be \label{eq:V_rho}
V(\xt) = \mathbb{P} \exp\left\{ i \int \ud x^-  
g^2 \frac{1}{ \nabt^2} \rho^a(\xt,x^-) \right\}, 
\ee
where $\mathbb{P}$ denotes path-ordering in $x^-$.
The color charge density is a random variable 
with a local  Gaussian probability distribution
\be \label{eq:S2}
P[\rho^a] \sim 
\exp\left\{- \int \ud^2 \xt \ud x^- \frac{\rho^a(\xt,x^-) \rho^a(\xt,x^-)}{2\mu^2(x^-)}\right\},
\ee
The total color charge $\int \ud x^- \mu^2(x^-) \sim \qs^2$ is
proportional to the thickness of a given
nucleus.

In the numerical calculation
the MV model initial conditions have been constructed as described in
Ref.~\cite{Lappi:2007ku}, discretizing the longitudinal coordinate $Y$
in $N_y=100$ steps.  
 For the calculations using the MV model directly 
for the initial conditions \nr{eq:alphai}, \nr{eq:alpha1+alpha2}
we have performed simulations on lattices of two different
sizes: $N_\perp=1024$, with the MV model color charge parameter $g^2
\mu L= 156$ which translates into $\qs a =0.119$; and with
$N_\perp=2048$, using $g^2 \mu L= 550$, which results in $\qs a =
0.172$.

The MV model also provides the configurations used as the initial
condition for quantum evolution in rapidity via the JIMWLK
renormalization group equation, starting at $Y=\log
x_0/x=0$. Performing a step $\Delta Y$ in rapidity opens phase space
for radiation of additional gluons which modify the classical
action~\nr{eq:V_rho}, \nr{eq:S2}. This process can be expressed as a
``random walk'' in the space of light-like Wilson lines
$V(\xt)$~\cite{Weigert:2000gi,*Mueller:2001uk,Blaizot:2002xy,Lappi:2012vw}:
\bea
&&\partial_Y V(\xt) =
V(\xt)
\frac{i}{\pi}\int \ud^2\ut
 \frac{(\xt-\ut)^i\eta^i(\ut)}{(\xt-\ut)^2}
\\ \nonumber
&&
- \frac{i}{\pi}\int \ud^2\vt 
  V(\vt) \frac{(\xt-\vt)^i \eta^i(\vt)}{(\xt-\vt)^2} V^\dag(\vt) 
V(\xt),
\label{eq:Lgvn}
\eea
where the Gaussian white noise $\eta^i =\eta^i_a t^a$ satisfies
$\langle \eta^a_i(\xt)\rangle =0$ and, for fixed coupling,
\be \label{eq:etaeta}
\langle \eta^a_i(\xt)\; \eta^b_j(\yt)\rangle = \as \delta^{ab}
\delta_{ij}\delta^{(2)}(\xt-\yt).
\ee
Here the equation is written in the left-right symmetric
form introduced in~\cite{Kovner:2005jc,Lappi:2012vw}.

The fixed coupling JIMWLK equation is solved using the numerical
method developed in \cite{Blaizot:2002xy,Rummukainen:2003ns,Lappi:2012vw}.
For the smaller lattice size $N_\perp=1024$ we start with the MV model
with $g^2\mu L = 31$ and without a mass regulator, which corresponds
to an initial $\qs a = 0.0218$. After $\Delta y = 1.68/\as$ units of
evolution in rapidity\footnote{For fixed coupling the evolution
  variable is $\as y$, so we do not need to specify a particular value
  of $\as$ separately.} this results in $\qs a = 0.145$. For a
$N_\perp=2048$-lattice we again start with $g^2\mu L = 31$,
corresponding to $\qs a = 0.0107$, and after $\Delta y = 1.8/\as$
units of evolution end up with $\qs a = 0.141$.

For running coupling the evolution is significantly slower. We use the
running coupling prescription introduced in \cite{Lappi:2012vw},
where the scale of the coupling is taken as the 
momentum conjugate to the distance in the noise correlator
in \eq\nr{eq:etaeta}. For
the smaller $N_\perp=1024$ lattice we again start with $g^2\mu L =
31$, i.e.  $\qs a = 0.0218$ and evolve for $\Delta Y = 10$ units in
rapidity, arriving at $\qs a = 0.118$.  For the larger $N_\perp=2048$
lattice we test a configuration that is farther from the IR cutoff,
starting the JIMWLK evolution with $g^2\mu L = 102.4$, i.e.\ $\qs a =
0.0423$ and evolve for $\Delta Y = 10$ units in rapidity, arriving at
$\qs a = 0.172$. In the rc-JIMWLK simulations the QCD scale is taken
as $\lqcd a = 0.00293$ and the coupling is  
frozen to a value $\alpha_0 = 0.76$ in the infrared below $2.5\lqcd$.

As already mentioned above, RG evolution in rapidity resums quantum
corrections to the fields $\alpha_m^\mu$ of the individual charge
sheets to all orders in $\alpha_s \log 1/x$, with leading logarithmic
accuracy. In other words, the effective action at $Y$ is modified from
that at $Y=0$, written in \eq\nr{eq:S2}.

Once an ensemble of Wilson lines $V(\xt)$ at a rapidity $Y$ is
constructed, separately  for both projectile and target,
these configurations define $\alpha_1^i$ and
$\alpha_2^i$ in light-cone gauge as written in \eq\nr{eq:alphai};
the initial field $A^i$ of produced soft gluons at proper time
$\tau=+0$ corresponds to their sum, \eq\nr{eq:alpha1+alpha2}.  The
evolution to $\tau>0$ follows from the real-time Hamiltonian evolution
described in Ref.~\cite{Krasnitz:1998ns}. This has been used in many
classical field calculations, e.g.\ in
Refs.~\cite{Krasnitz:2001qu,*Krasnitz:2003jw,*Lappi:2003bi}, or more
recently for the first study of the effects of JIMWLK-evolution on the
gluon spectrum~\cite{Lappi:2011ju}, and in the IP-glasma model for the
initial conditions for hydrodynamics~\cite{Schenke:2012hg}.
 On the $N_\perp=2048$
lattices we evolve the fields up to $\qs \tau=5$ and on the smaller
$N_\perp=1024$ ones to $\qs \tau=10$. In this study, the nuclei are 
taken to fill the whole transverse lattice, with periodic boundary conditions.

%%%%%%%%%%%%%%%%%%%%%%%%%
\section{Wilson loop}
\label{sec:wloop}
%%%%%%%%%%%%%%%%%%%%%%%%%%

\begin{figure}[t!]
\centerline{\includegraphics[width=0.45\textwidth]{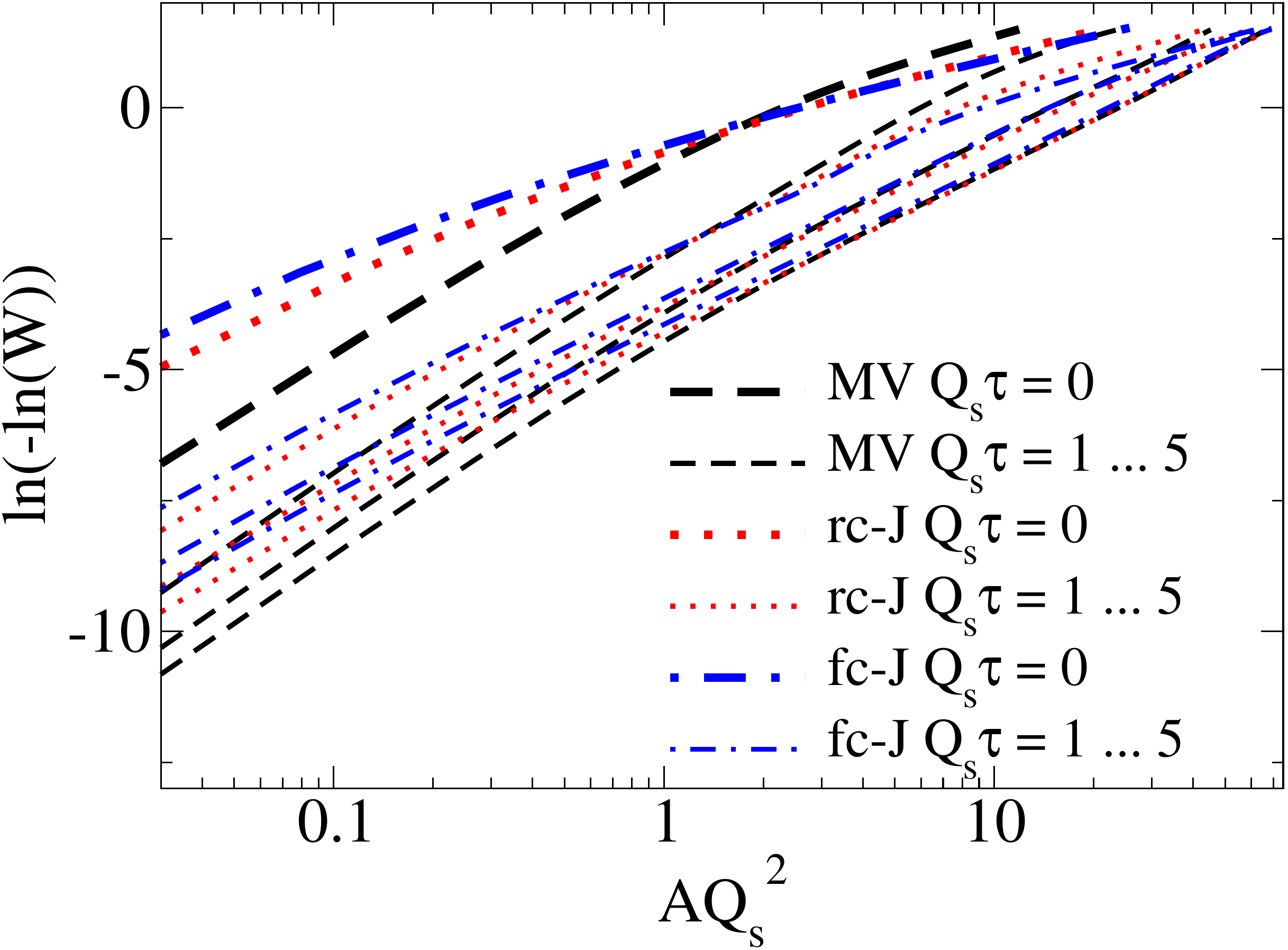}}
\caption{Wilson loop as a function of area for different initial
  conditions and times measured on $N_\perp=2048$ lattices. The
  thicker lines at the top correspond to time $\qs \tau=0$, for the
  classical MV model as well as for fixed and running coupling JIMWLK
  evolution. The results for $\qs \tau=1,3,5$ are shown by the thinner
  lines, with later times corresponding to smaller values of $\ln(-\ln
  W)$.
\label{fig:fitchk}
}
\end{figure}

In the continuum the spatial (magnetic) Wilson loop is defined as the
trace of a path ordered exponential of the gauge field around a closed
path of area $A$ in the transverse plane:
\begin{equation}
W(A) = \frac{1}{\nc} \left< \tr \mathbb{P} \exp\left\{i g 
\oint_{\partial A} \ud \xt \cdot \At \right\} \right>. 
\end{equation}
On the lattice this is easily discretized as the product of 
link matrices around a square of area $A$. For $N_c\ge3$ colors any
particular Wilson loop is complex but the ensemble average is real.

We have measured the expectation value of the Wilson loop in the
glasma field, with different initial conditions and at different times
$\qs \tau$. The results of the calculation are shown in
\fig\ref{fig:fitchk}. As expected, the magnetic flux through a loop
generically increases with its area. Focusing first on the curves
corresponding to the initial time $\tau=0$ we observe that the
resummation of quantum fluctuations (JIMWLK evolution) increases the
flux through small loops of area $A\qs^2<1$. This can be understood
intuitively as due to emission of additional virtual soft gluons in
the pure gauge fields of the colliding charge sheets. On the other
hand, the flux through large loops, $A\qs^2\gsim 2$, decreases. This
indicates uncorrelated fluctuations of magnetic flux over such areas
and is consistent with the suggestion that the flux is ``bundled'' in
domains with a typical area $\sim1/\qs^2$~\cite{Dumitru:2013koh}. 
Accordingly, loops of area $\sim 1.5\qs^2$ are invariant under
high-energy evolution.

Moving on to finite times we see that the flux through loops of fixed
area decreases with $\tau$. This is, of course, a consequence of the
decreasing field strength in an expanding metric. For small loops the
ordering corresponding to the different initial conditions (MV,
rc-JIMWLK, fc-JIMWLK) is preserved even at later times. However, for
large loops one observes a striking ``universality'' emerging at
$\qs\tau\sim5$ as the curves for all initial conditions fall on top of
each other.

\begin{figure}[t]
\centerline{\includegraphics[width=0.45\textwidth]{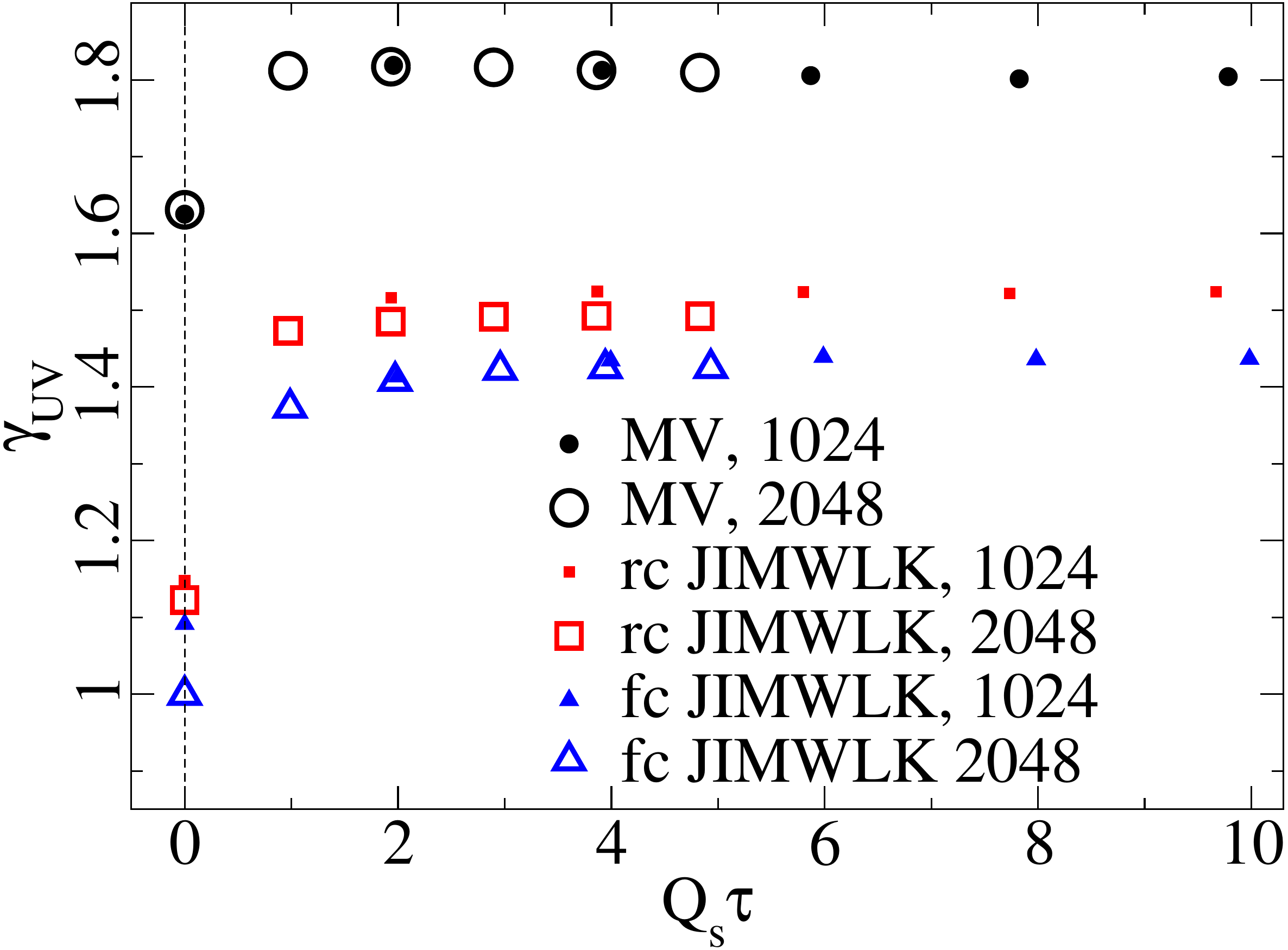}}
\caption{
Time dependence of the exponents $\gamma$ in the parametrization 
\nr{eq:fitform} fitted to the UV region.
} \label{fig:uvexpvstau}
\end{figure}

\begin{figure}[t]
\centerline{\includegraphics[width=0.45\textwidth]{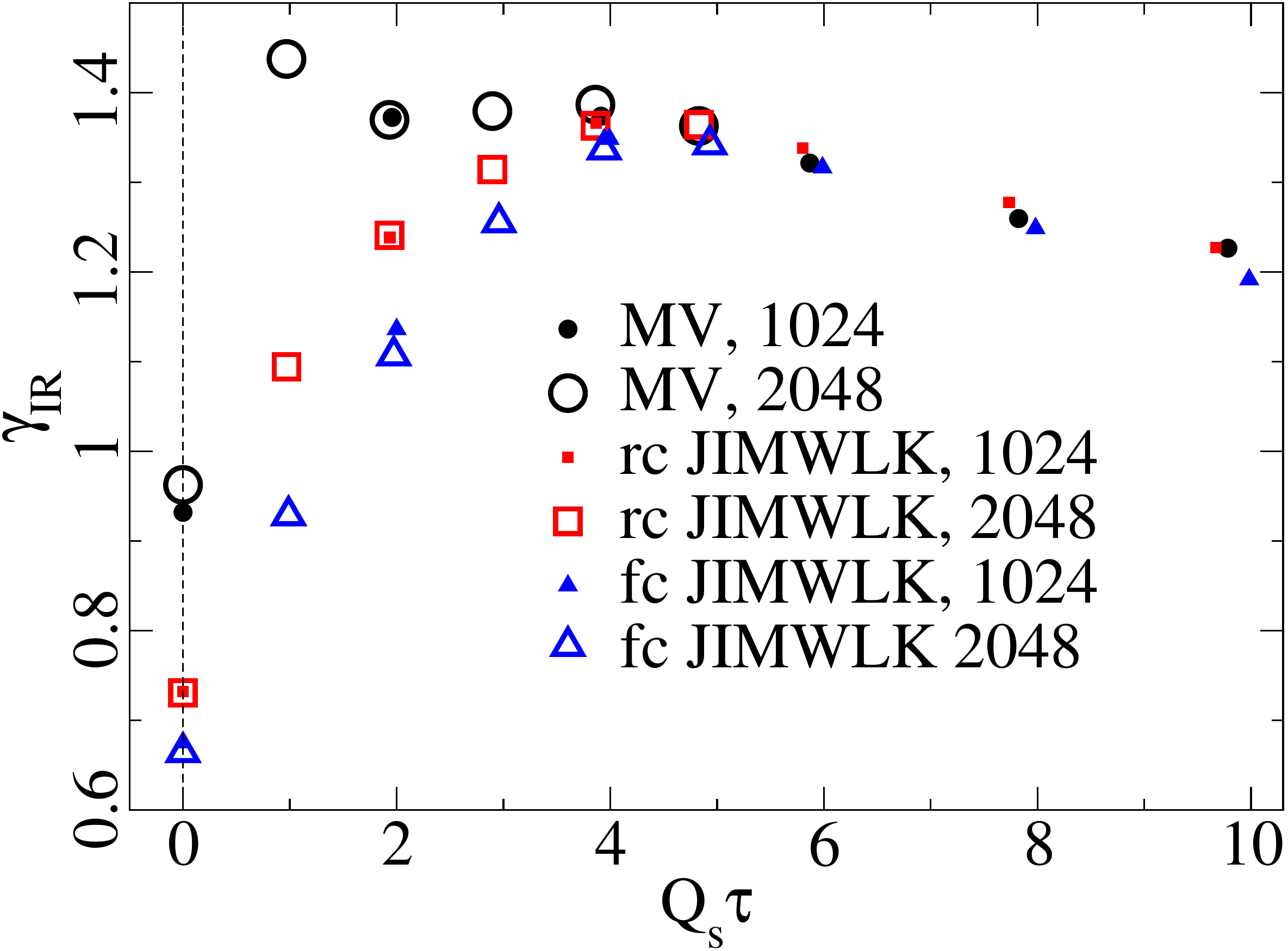}}
\caption{
Time dependence of the exponents $\gamma$ in the parametrization 
\nr{eq:fitform} fitted to the IR region.
} \label{fig:irexpvstau}
\end{figure}

\begin{figure}[t!]
\centerline{\includegraphics[width=0.45\textwidth]{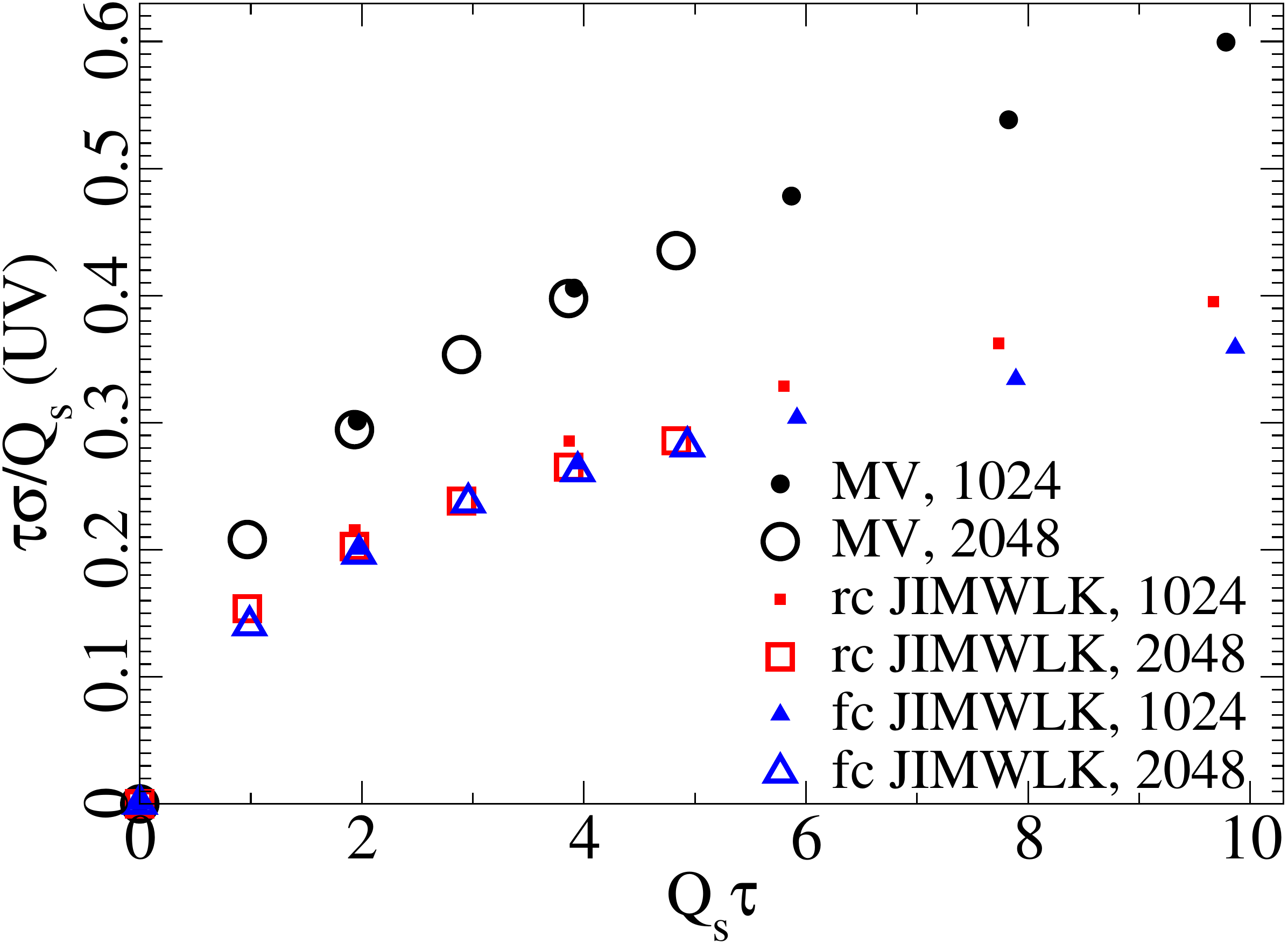}}
\caption{Time dependence of ``string tension'' coefficient $\sigma$
  fitted to the UV region.  The values of $\sigma/\qs^2$ at $\tau=0$
  are 0.59 [0.57]; 0.55 [0.53] and 0.56 [0.56] for the MV, rcJIMWLK
  and fcJIMWLK initial conditions respectively on a $N_\perp=1024$
  [$N_\perp=2048$] lattice.  } \label{fig:uvsigmavstau}
\end{figure}

\begin{figure}[t!]
\centerline{\includegraphics[width=0.45\textwidth]{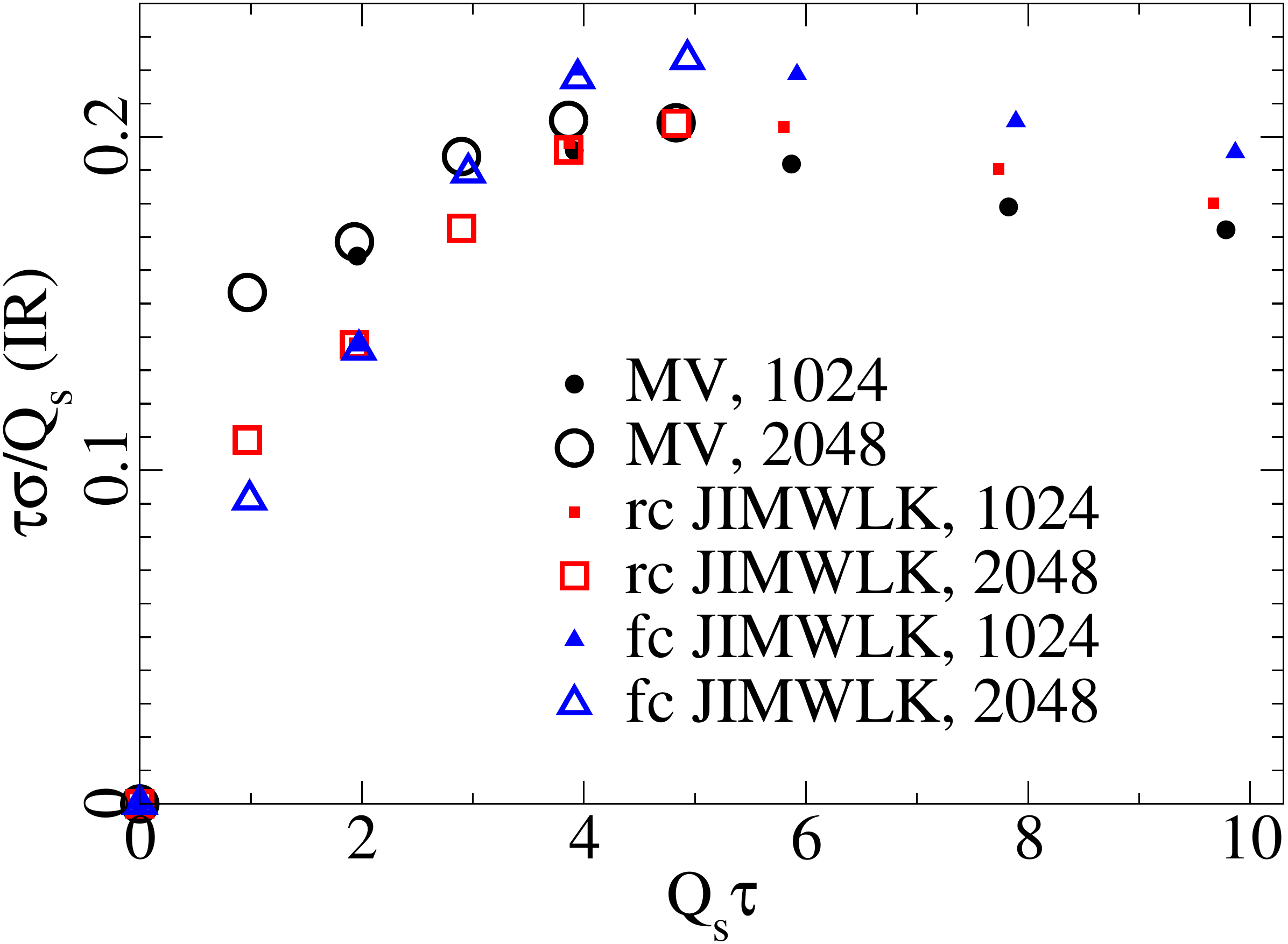}}
\caption{Time dependence of ``string tension'' coefficient $\sigma$
  fitted to the IR region, multiplied by $\tau$ to separate out the
  natural $\sigma \sim 1/\tau$ dependence due to the expansion of the
  system.  The values of $\sigma/\qs^2$ at $\tau=0$ are 0.43 [0.44];
  0.37 [0.38] and 0.39 [0.40] for the MV, rcJIMWLK and fcJIMWLK
  initial conditions respectively on a $N_\perp=1024$ [$N_\perp=2048$]
  lattice.  } \label{fig:irsigmavstau}
\end{figure}

The data from \fig\ref{fig:fitchk} shows an approximately linear
dependence of $\ln(-\ln W)$ on $\ln(A\qs^2)$, with different slopes
in the regime of small $A\qs^2\ll 1$ vs.\ large $A\qs^2\gg 1$. Based on
this observation we fit the data to 
\begin{equation}\label{eq:fitform}
W(A) = \exp\left\{-(\sigma A )^\gamma \right\},
\end{equation}
with separate parametrizations for the IR and UV regimes:
\bea
\mathrm{IR:} && e^{0.5} < A \qs^2 < e^{5} \\
\mathrm{UV:} && e^{-3.5} < A \qs^2 < e^{-0.5}~.
\eea
In addition to limiting the fits to the quoted ranges we also restrict
them to the region where $W>0.01$ and the statistical error
on $W$ is less than 0.2$W$; beyond these limits the data exhibits too large
fluctuations for a meaningful fit. Figures \ref{fig:uvexpvstau} and
\ref{fig:irexpvstau} show the time dependence of the exponents
$\gamma$ in the IR and UV regions. The ``string tension'' $\sigma$
naturally decreases as $\sim 1/\tau$ because of the longitudinal
expansion of the glasma, which leads to $B_z \sim 1/\sqrt{\tau}$. We
therefore show, in \figs\ref{fig:uvsigmavstau}
and~\ref{fig:irsigmavstau}, the time dependence of the combination
$\tau\sigma/\qs$, where this leading effect is scaled out. The values
of $\sigma/\qs^2$ for $\qs \tau = 0$ are given in the
captions\footnote{For the MV model, at $\tau=0$ we find
  $\sigma/\qs^2=0.44$ in the IR region which is about four times
  larger than the value reported in ref.~\cite{Dumitru:2013koh}. Our
  present results refer to $N_c=3$ colors while
  ref.~\cite{Dumitru:2013koh} considered $N_c=2$; also, our current
  definition of $\qs$ via \eq\nr{eq:defqs} leads to
  smaller values for this quantity than the definition used
  in~\cite{Dumitru:2013koh}. Finally, $\sigma$ is extracted from fits
  over a somewhat different range of areas.}.


The results in the ``UV''-regime probed by Wilson loops of small area are
shown in \figs\ref{fig:uvexpvstau} and~\ref{fig:uvsigmavstau}. They are
easily understood from the differences in the initial condition. The
gluon spectrum in the MV model falls steeply as a function of $\ktt$,
leading to a steep dependence of short-distance correlators on the
distance. Our result for the UV exponent in \fig\ref{fig:uvexpvstau}
is close to the $A^2$-scaling obtained analytically in a weak field
expansion~\cite{Petreska:2013bna}. The difference is probably due to a
combination of logarithmic corrections and lattice UV cutoff
effects. For the JIMWLK ensembles the gluon spectrum is much
harder~\cite{Lappi:2011ju}, especially for fixed coupling. This
manifests itself in smaller values of both $\gamma$ and $\sigma$. In
addition, the UV exponents are remarkably time independent at
$\qs\tau>1$: this is consistent with the expectation that at such time
the UV modes can be viewed as noninteracting gluons whose spectrum is
close to the expectation from a perturbative $\ktt$-factorized
calculation~\cite{Blaizot:2010kh}.

The behavior in the IR regime (\figs\ref{fig:irexpvstau},
\ref{fig:irsigmavstau}) probed by large Wilson loops points to a
very different picture. At $\tau=0$ the exponents $\gamma$ and, to a
lesser extent, the values of $\sigma$ depend very much on the initial
conditions. As already alluded to above, the scaling exponents
$\gamma_{\rm IR}<1$ obtained for the JIMWLK fields indicate that
quantum emissions increase magnetic flux fluctuations at the scale
$\sim1/\qs$, much smaller than the area of the loop. It is interesting to
note that for the rather strong fixed-coupling evolution the initial
scaling exponent is not too far above $\gamma_{\rm IR}=1/2$ corresponding
to perimeter scaling.

At times $\qs\tau \gtrsim 3$, however, one observes a remarkable
universality in the IR as the curves corresponding to different initial
conditions collapse onto a single curve in \fig\ref{fig:fitchk}. 
The string tensions in \fig\ref{fig:irsigmavstau} are within 10\% of
each other at late $\qs\tau$, and the exponents $\gamma$ in
\fig\ref{fig:irexpvstau} are very  close to each other, with
values around $\gamma_{\rm IR}\approx 1.2 \dots 1.3$. The exponent gradually
decreases with $\tau$, potentially approaching the area law $\gamma=1$
at late times. 
The initial evolution
points at a rapid rearrangement of ``magnetic hot spots'' to some
universal field configurations at later time, $\qs\tau \gtrsim 3$.

We stress that the \emph{universal} behavior of
large magnetic loops, characterized by a nontrivial power-law dependence
on the loop area, sets in at rather early time scales of a few times
$1/\qs$, independent of initial conditions.
Actual area law  scaling $\gamma=1$ is approached only later. 
This behavior
mirrors a similar universality between MV and JIMWLK results seen in
the IR part of the gluon spectrum (determined from correlators of
gauge fixed fields) in Ref.~\cite{Lappi:2011ju}. 
Since the structure of the fields does not seem to depend on the initial 
conditions, we infer that this universality in due to stong interactions
in the glasma phase.
This universal behavior of
the Wilson loop for different initial conditions at $\qs\tau \gtrsim
3$ and $A\qs^2 \gg 1$ is the main result of this paper.

%%%%%%%%%%%%%%%%%%%%%%%%%
\section{Magnetic field correlator}
\label{sec:bbcorr}
%%%%%%%%%%%%%%%%%%%%%%%%%%

In this section we analyze gauge-invariant
two-point magnetic field correlators of the
form\footnote{We include the factor $g^2$ for convenience, because the
quantity that appears naturally in the classical lattice formulation
is  actually $gB$.}
\begin{equation}
C_B(r) \equiv 2 g^2 \tr  \left< B_z(\xt) U_{\xt\to \yt}
B_z(\yt) U^\dag_{\xt\to \yt} \right>~.
\end{equation}
The points $\xt$ and $\yt$ are separated in the $x$ or $y$
direction by a distance $r=|\xt-\yt|$, and the Wilson line
$U_{\xt\to \yt}$ is the ordered product of links along the straight
line separating these points.

The magnetic field $B_z = t^a B_z^a$ 
on the lattice is defined as the traceless
antihermitian part of the plaquette as
\begin{equation}
g B_z^a (\xt) = 2 \, \mathrm{Re} \tr t^a U_{x,y}(\xt)~,
\end{equation}
where the transverse plaquette is
\begin{equation}
U_{i,j}(\xt) = U_i(\xt) U_j(\xt+\itt) U^\dag_i(\xt+\jt) U^\dag_j(\xt)~.
\end{equation}
Here $U_i(\xt)$ denotes the link matrix in the $i$-direction based at $\xt$
and $\itt,\jt$ are unit vectors.

\begin{figure}[t!]
\centerline{\includegraphics[width=0.45\textwidth]{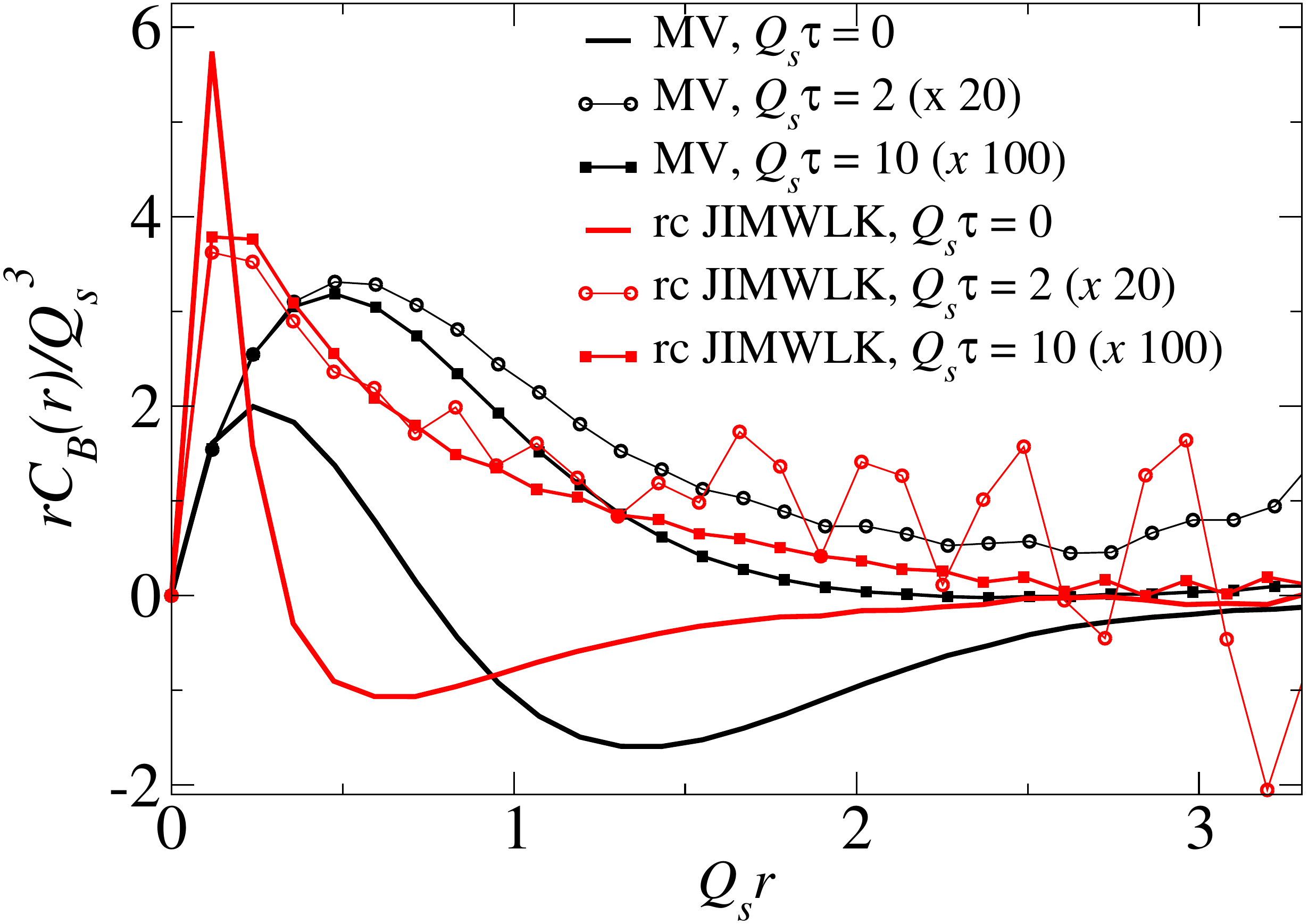}}
\caption{Magnetic field correlator on a $1024^2$-lattice at $\qs \tau
  =0$, 2, and 10; the latter have been rescaled by factors of 20 and
  100, respectively.} \label{fig:Bcorr}
\end{figure}
The resulting magnetic field correlator $r\; C_B(r)$ is plotted in
\fig\ref{fig:Bcorr}. We have multiplied by $r$ to better expose the
behavior around $r \sim 1/\qs$.  At the initial time there is a
significant anticorrelation at intermediate distances. It shows the
domain structure of the field such that $B_z$ is likely to flip
sign\footnote{Recall that $B$ transforms homogeneously. Hence, unlike
  the links $U_i(\xt)$, the magnetic field $B_z(\xt)$ can be
  diagonalized everywhere by a suitable gauge transformation.} (or
direction) over distances of order $1/\qs$.  This structure then
changes very rapidly: already at time $\qs \tau \sim 2$ the fields
have rearranged such that the anticorrelation has disappeared. Also,
the subsequent time evolution results in damping of the fluctuations
at $\qs r \gsim 1$ which are present in the initial field
configurations. On the other hand, the strong short-distance
correlations around the peak are not affected much by the time
evolution beyond $\qs \tau \sim 2$, aside from a decrease in
magnitude. In particular, no ``infrared diffusion'' of the peak
towards larger distances is observed.

\begin{figure}[t!]
\centerline{\includegraphics[width=0.45\textwidth]{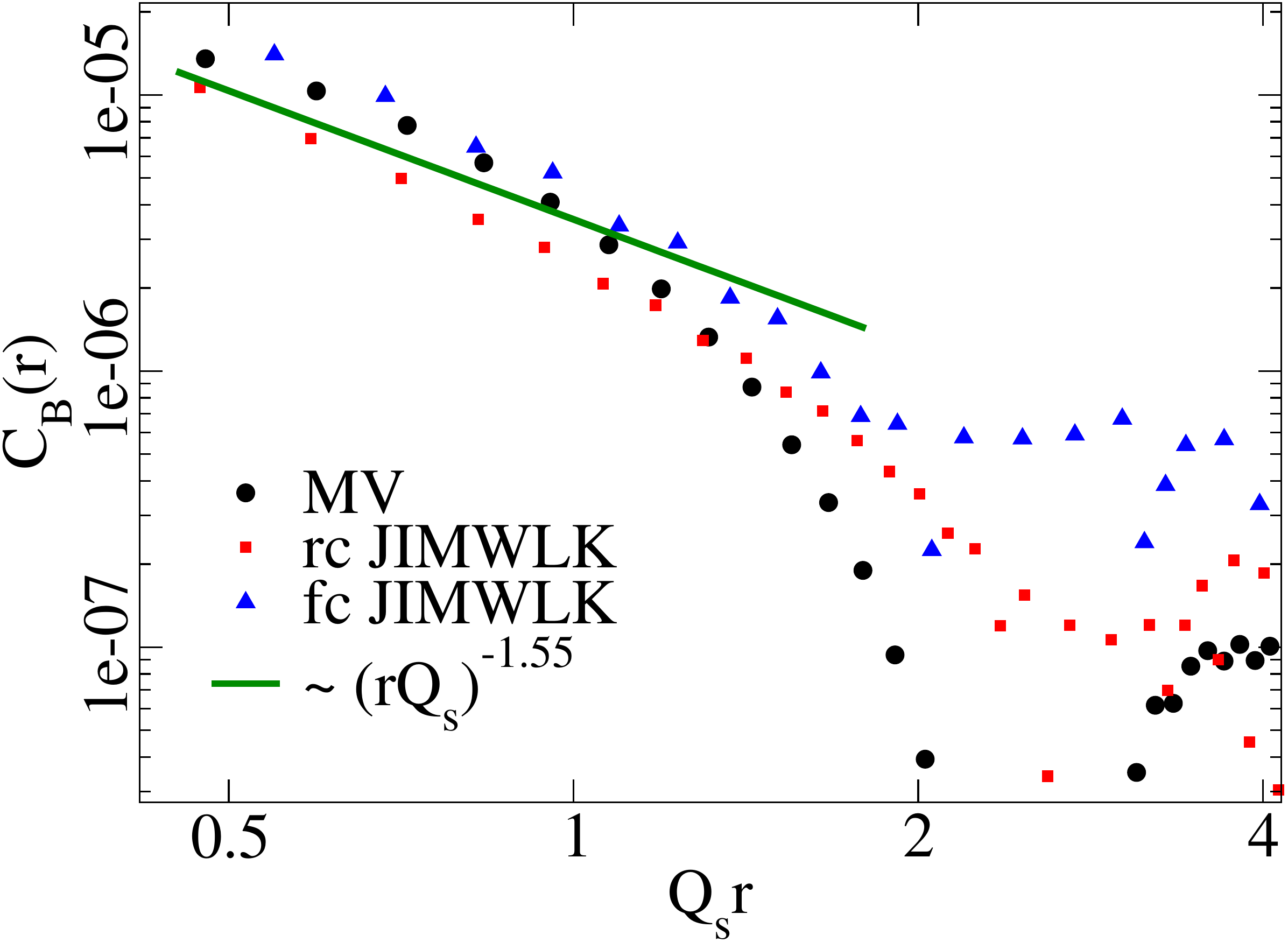}}
\caption{ Magnetic field correlator $C_B(r)$ at $\qs \tau=10$ on a
  $1024^2$-lattice.  The line corresponds to $\sim r^{-\alpha}$ with
  the exponent $\alpha= 4-2\gamma_\textnormal{IR}=1.55$ extracted in
  the previous section from
  the fit of $\gamma_\textnormal{IR}$ to the Wilson loop.} \label{fig:bbscaling}
\end{figure}
Given the clear scaling behavior of the Wilson loop one might expect to 
see a similar phenomenon for the magnetic field correlator. A very naive 
scaling argument would assume that if $C(r) \sim r^{-\alpha}$ then
the Wilson loop should scale as
\begin{multline}
- \ln W \sim \int\limits_A \ud^2 \xt \ud^2 \yt \, C(|\xt-\yt|) 
\\ \sim  
R^{4-\alpha} \sim A^{\frac{4-\alpha}{2}} = A^{\gamma}~.
\end{multline}
The area integrals in the first line extend over $|\xt|,|\yt|<R$.
Thus, $\gamma = 1.225$ extracted from the Wilson loop at $\qs \tau=10$
would give $C_B(r) \sim r^{-1.55}$. On a logarithmic scale $C_B(r)$
does indeed qualitatively resemble such behavior as shown in
\fig\ref{fig:bbscaling}. However, this kind of scaling is less
conclusive than for the Wilson loop (see  also appendix); this could 
be an indication for the presence of higher cumulants in the expansion of
the spatial Wilson loop~\cite{Dosch:1988ha,*DiGiacomo:2000va}.

\begin{figure}[t!]
\centerline{\includegraphics[width=0.45\textwidth]{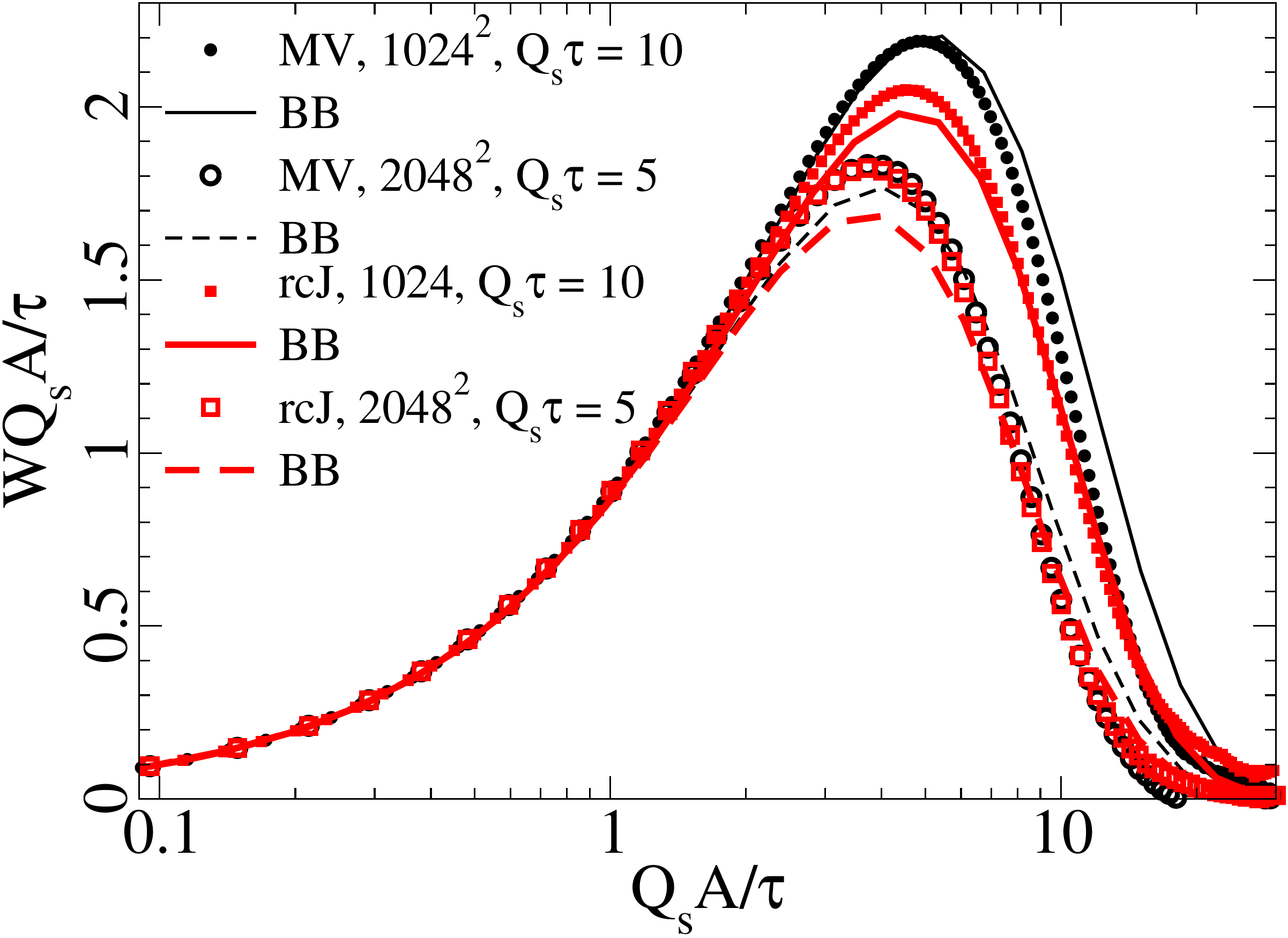}}
\caption{Direct measurement of the Wilson loop (points) compared to an
  approximation in terms of the Gaussian cumulant,
  \eq\nr{eq:wlfrombb}, which reconstructs it from the magnetic field
  correlator (lines).  } \label{fig:wlscfrombb}
\end{figure}

%%%%%%%%%%%%%%%%%%%%%%%%%
\section{Summary}
%%%%%%%%%%%%%%%%%%%%%%%%%

In this paper we have provided some insight into the fields produced
initially in a high-energy collision of dense color charge sheets. We
have focused, in particular, on the structure of the longitudinal
magnetic field $B_z$.

We consider both purely classical as well as JIMWLK RG evolved gauge
field ensembles on which we measure expectation values of spatial
Wilson loops and two-point correlation functions of $B_z$.  These show
that the initial fields exhibit domain-like structure over distance
scales of the order of the inverse saturation scale $1/\qs$. Classical
YM evolution to later times leads to universal scaling, for all
ensembles, of the magnetic loop with area, with a nontrivial critical
exponent. Also, the anti-correlation of $B_z(\xt)$ over distances
$\sim 1/\qs$ disappears, which we interpret as rearrangement, possibly
accompanied by transverse expansion, of the magnetic field domains.

The emergence of a color field condensate in high-energy collisions of
dense hadrons or nuclei is a very interesting phenomenon, and its
dynamics remains to be understood in more detail. In closing we only
draw attention to recent arguments that the presence of such a
condensate might have important implications for the process of (pre-)
thermalization in high multiplicity collisions~\cite{Floerchinger:2013kca}.

%%%%%%%%%%%%%%%%%%%%%%%%%
\appendix
\section{Relation between the Wilson loop and magnetic field correlator}
%%%%%%%%%%%%%%%%%%%%%%%%%

In an abelian theory there is a simple relation between the Wilson 
loop and the magnetic field due to Stokes' theorem:
\begin{equation}
\oint_{\partial A} \ud \x \cdot \Avec =
\int_A \ud^2 \x \, B_z(\x) ~.
\end{equation}
If we assume that in the nonabelian case the magnetic field in each
color channel $a$ is independent, and that it consists of uncorrelated
domains which are much smaller than the area $A$ and distributed as Gaussian
random variables, we obtain the following estimate for the Wilson loop:
\begin{multline}\label{eq:wlfrombb}
\frac{1}{\nc} \tr \exp\left\{i g 
\oint_{\partial A} \ud \x \cdot \Avec \right\} 
\\ 
\approx   \exp\left\{- \frac{g^2}{2\nc} \left<  \tr 
\left[\int_A \ud^2\x \, B_z(\x)\right]^2 \right>\right\}  
\\
= \exp\left\{-  \frac{1}{4\nc}  
\int_A \ud^2\x \ud^2\y \, C_B (\x-\y)\right\}. 
\end{multline}
In \fig\ref{fig:wlscfrombb} we compare the result of a numerical
integration of the r.h.s.\ of \eq\nr{eq:wlfrombb} using the measured
magnetic field correlator, to the direct measurement of the Wilson
loop. It can be seen that the two are in a relatively good
agreement. This consistency check supports the interpretation of $B_z$
as independent field domains of area $\sim 1/\qs^2$.

\section*{Acknowledgements}
T.~L.\ is supported by the Academy of Finland, projects 133005, 
267321 and 273464. This work was done using computing resources from
CSC -- IT Center for Science in Espoo, Finland.
A.~D.\ acknowledges support by the
DOE Office of Nuclear Physics through Grant No.\ DE-FG02-09ER41620 and
from The City University of New York through the PSC-CUNY Research
Award Program, grant 66514-0044.
The authors thank the Yukawa Institute for Theoretical Physics, 
Kyoto University, where part of this work was done 
during the YITP-T-13-05 workshop on ``New Frontiers in QCD''.

\bibliography{spires}
\bibliographystyle{JHEP-2modM}

\end{document}